\documentclass[preprint2]{aastex}
\def\imgsa{
\begin{figure*}
\epsscale{2}
\plotone{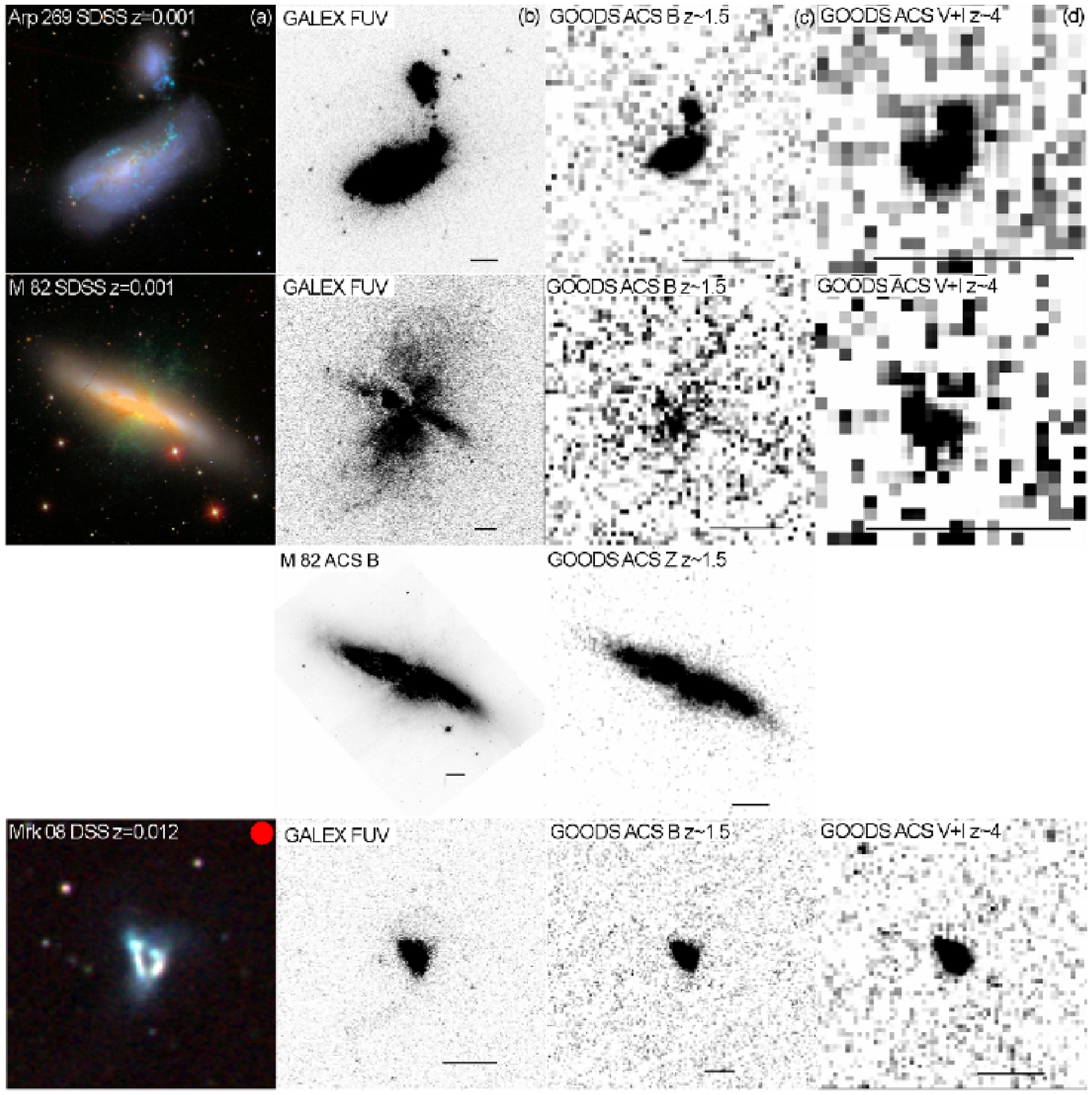}
\caption{Artificially redshifted sample. The columns are as follows (except in the case for M82 B-band): (a) color composite optical image from SDSS or DSS, as labeled, the original redshift is listed; (b) GALEX FUV image (ruler marks $1\arcmin$); (c) artificially redshifted to z$\sim1.5$ made to simulate GOODS \Bbnd\hspace{1pt} observations (ruler marks $1\arcsec$); (d) artificially redshifted to z$\sim4$ made to simulate GOODS \Vbnd+\ibnd\hspace{1pt} observations (ruler marks $1\arcsec$). The red dots indicate objects we determined to be analogs to high redshift clumpy star-forming galaxies (see \S \ref{analogs}).}\label{imgsa} 
\end{figure*}
}

\def\imgsb{
\begin{figure*}
\epsscale{2}
\plotone{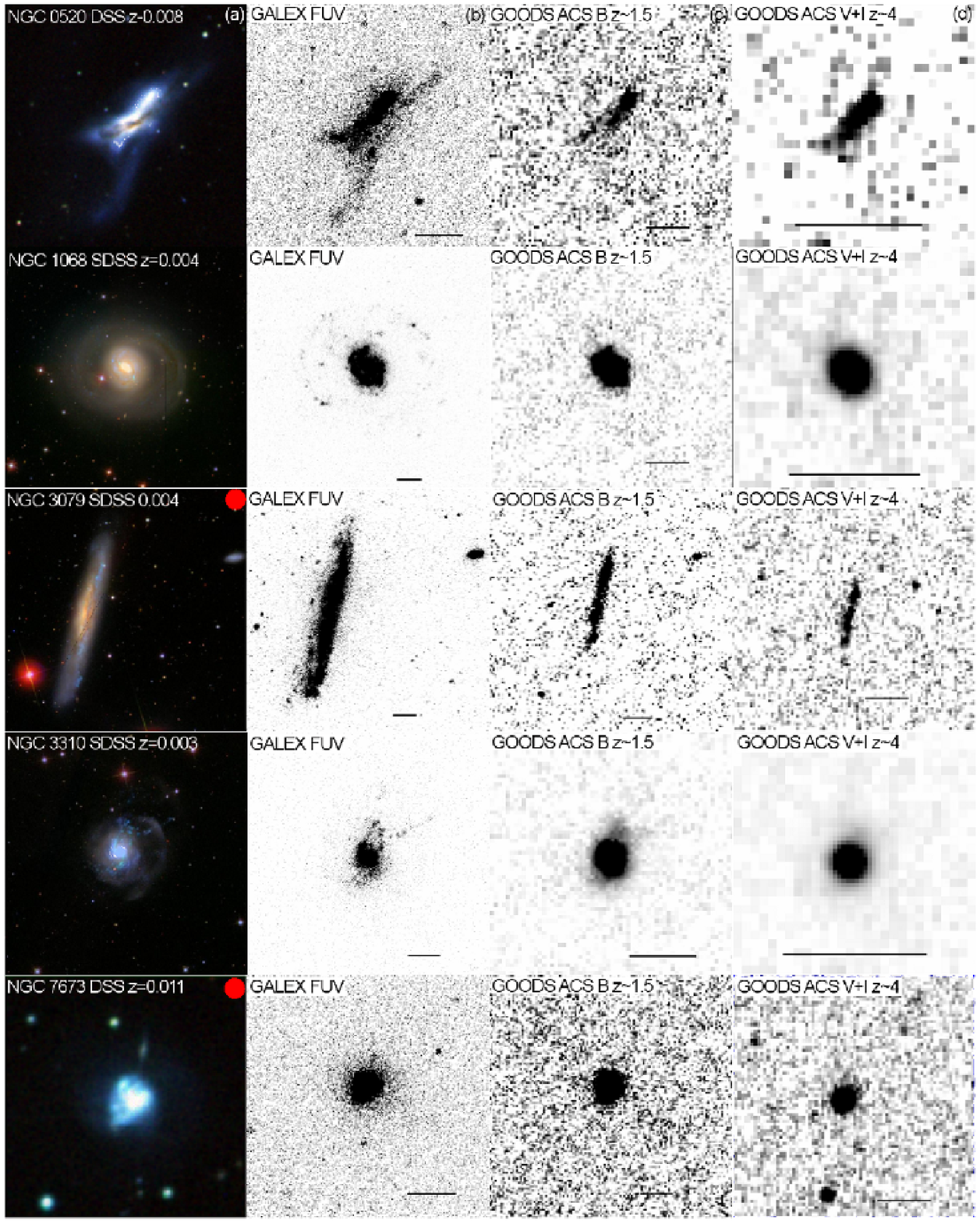}
\caption{Artificially redshifted sample. See Figure \ref{imgsa} for an explanation.}\label{imgsb}
\end{figure*}
}

\def\galfita{
\begin{figure*}
\epsscale{1.6}
\plotone{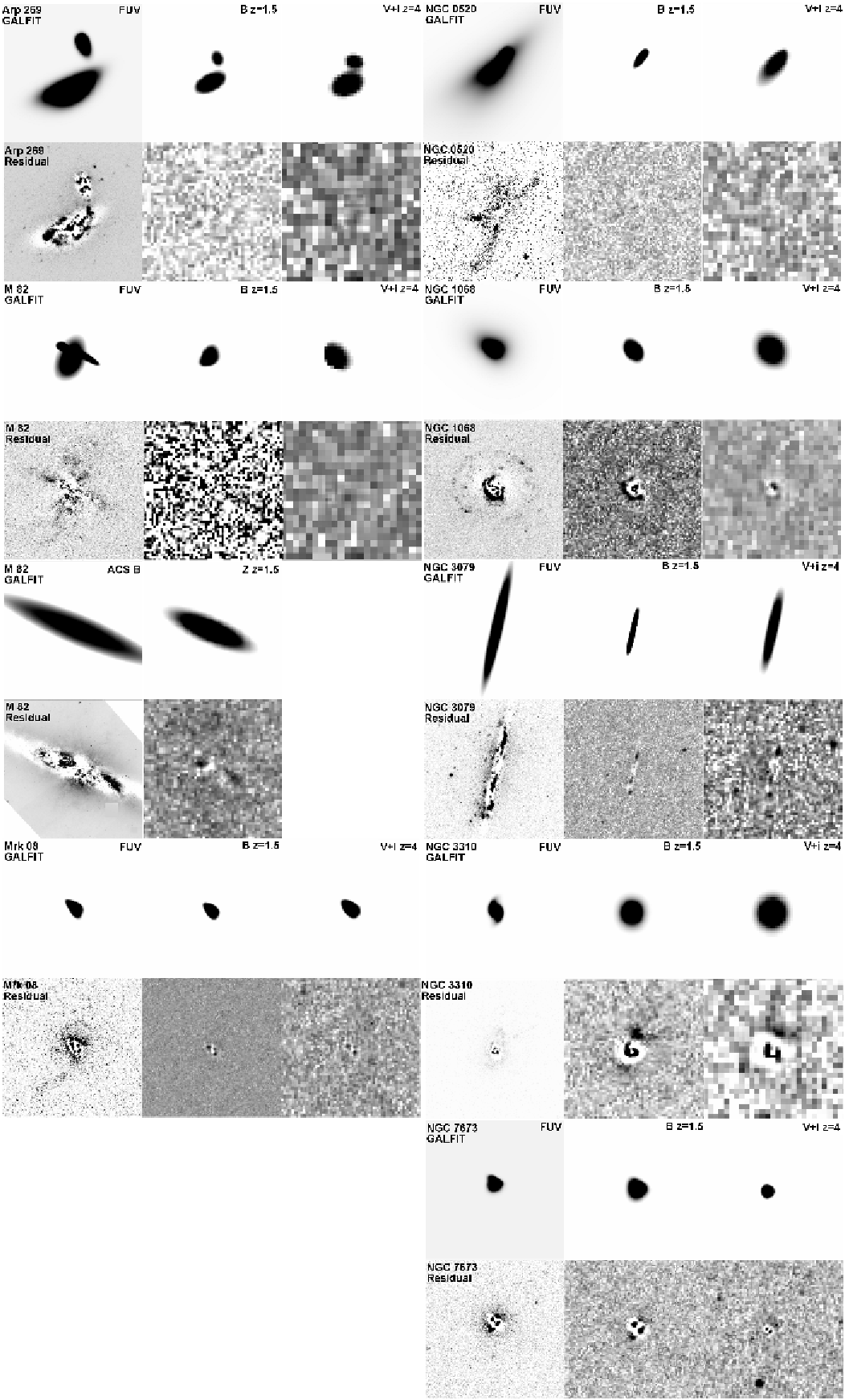}
\caption{GALFIT models for each artificially redshifted object. The GALEX FUV, GOODS \Bbnd, and GOODS \Vbnd+\ibnd\hspace{1pt} are labeled on the model fit image (M 82 ACS \Bbnd, and GOODS \zbnd \hspace{1pt} are labeled accordingly.). Each object has a top and bottom row, which show the 2D model (top) and the residual image (bottom).
}\label{galfita}
\end{figure*}
}

\def\comparison{
\begin{figure*}
\epsscale{2}
 \plotone{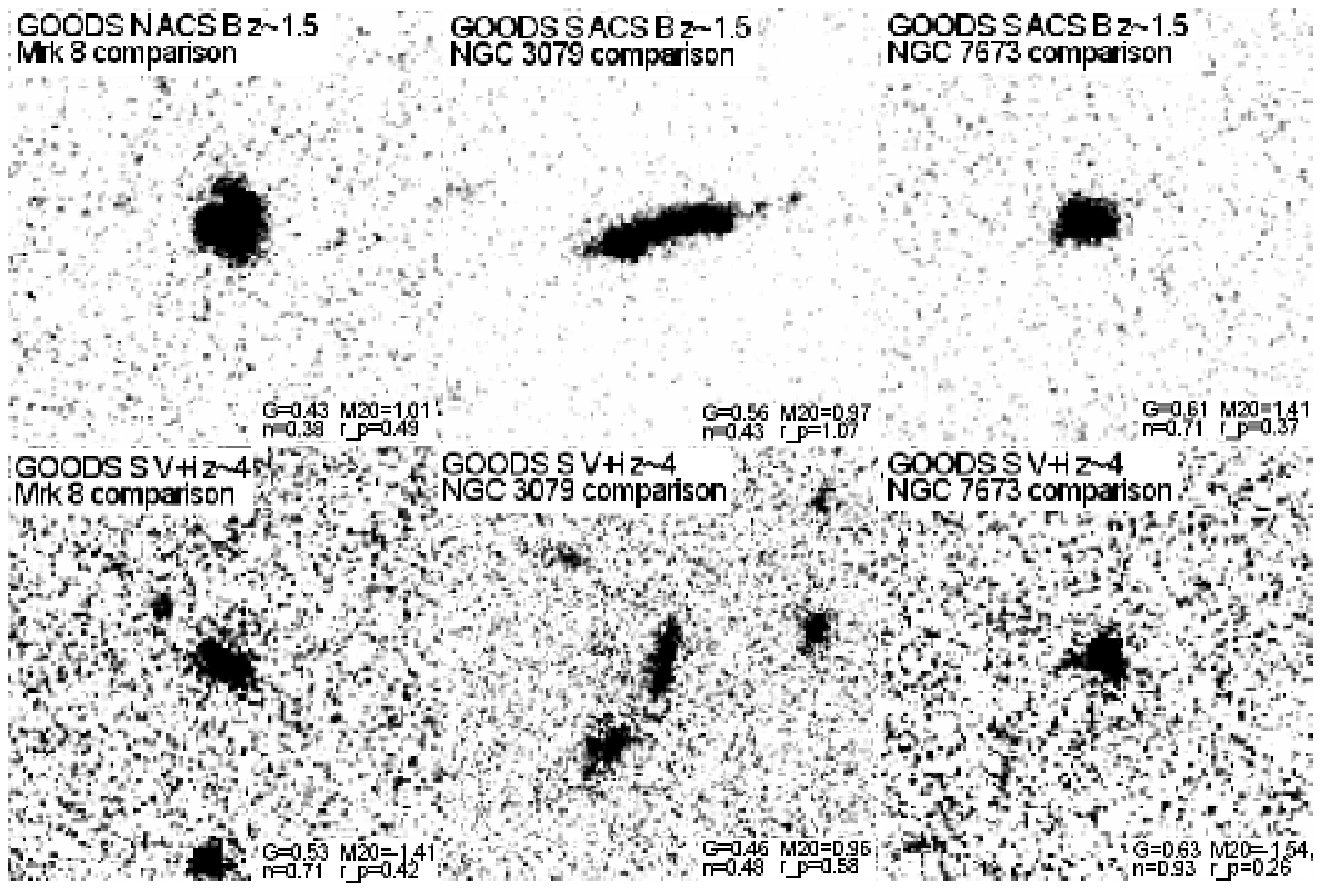}
\caption{Selected objects from the GOODS comparison sample that closely resemble Mrk 08, NGC 3079, and NGC 7673 at the simulated redshifts (see \S \ref{compgal}). The Gini, $M_{20}$, $n$, and $\rm{r\_p}$ values are listed on the bottom of each image. For z$\sim1.5$ ($\sim4$, B-dropouts) we used GOODS/ACS \Bbnd (\Vbnd+\ibnd) images. The sample was taken from \cite{lotz06} for all except the NGC 3079 z$\sim1.5$ comparison, which is from \cite{voy}.}\label{comparison}
\end{figure*}
}

\def\morphplot{
\begin{figure*}[h]
\epsscale{2.}
\centering
\plotone{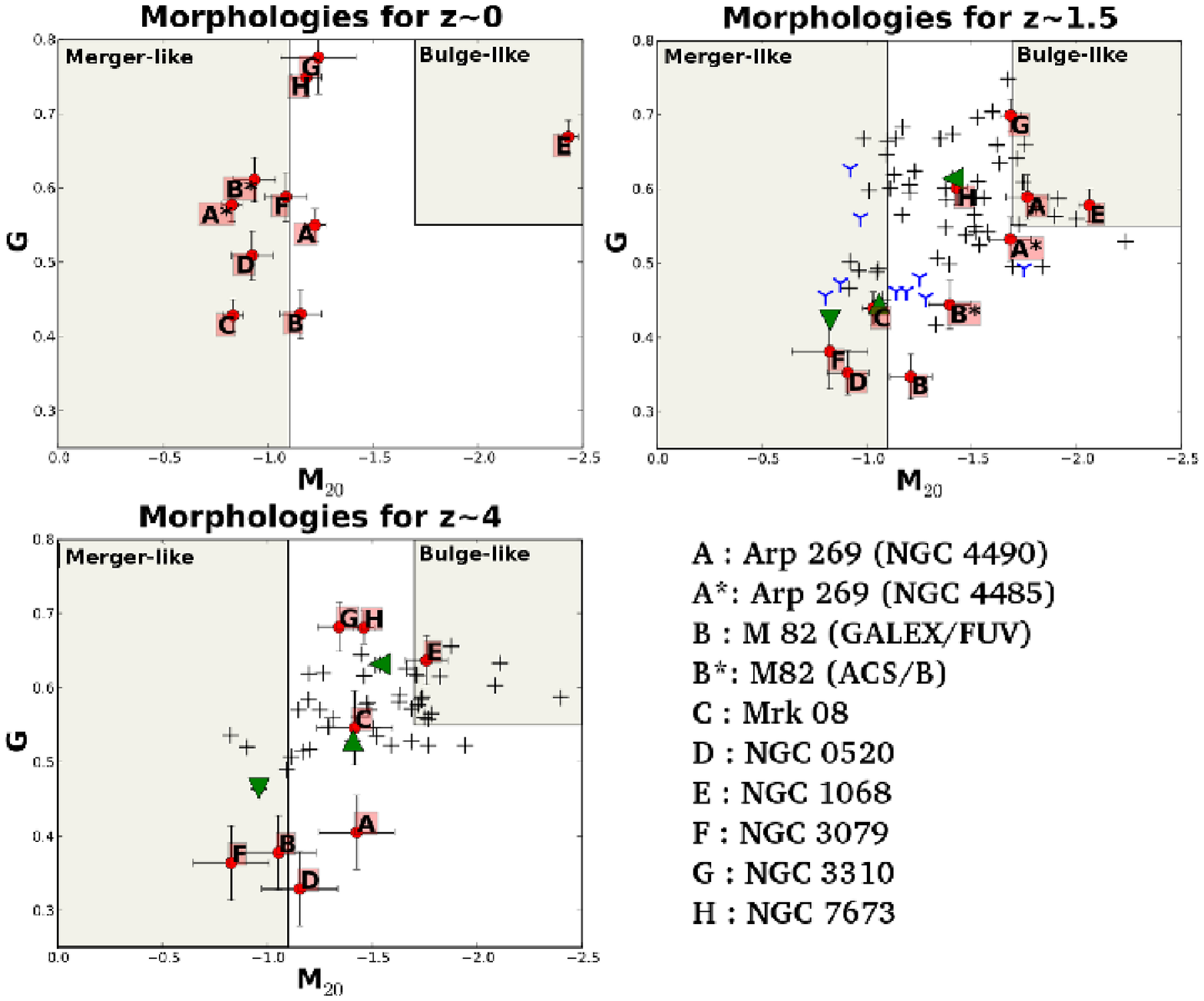}
\caption{Morphologies using Gini, and $M_{20}$. The red circles indicate the GALEX local universe sample artificially redshifted to z$\sim 1.5$ to 4. The black crosses are the GOODS/ACS sample from \cite{lotz06} in the \Bbnd-band for z$\sim1.5$ and \Vbnd+\ibnd-bands for z$\sim4$ LBGs. The blue tripods are GOODS/ACS \Bbnd-band objects from \cite{voy}. The up-, down- and left-pointing green triangles are z$\sim1.5$ and 4 starbursts that are morphologically similar to the Mrk 08, NGC 3079, and NGC 7673, respectively (see \S \ref{compgal}). The top left shows the morphologies for the original GALEX FUV images. The top right (bottom left) plot shows the morphologies for z $\sim 1.5$ ($4$). The shaded regions $0<$M$_{20}<-1.1$ and $-1.7<$M$_{20}<-2.5$ distinguish between merger- and bulge-dominated galaxies. Error bars are based on \cite{lotz06} Fig. 1. The original GALEX objects tend to be in the merger-dominated region. The z$\sim1.5$ and $4$ objects show a more diverse morphology as they shift into the intermediate region. We find that 20\% (11/54) of z$\sim1.5$ and 37\% (17/46) of z$\sim4$ galaxies are bulge-like.}\label{morphplot}
\end{figure*}
}

\def\morpha{
\begin{figure*}[h]
\centering
\epsscale{2.}
\plotone{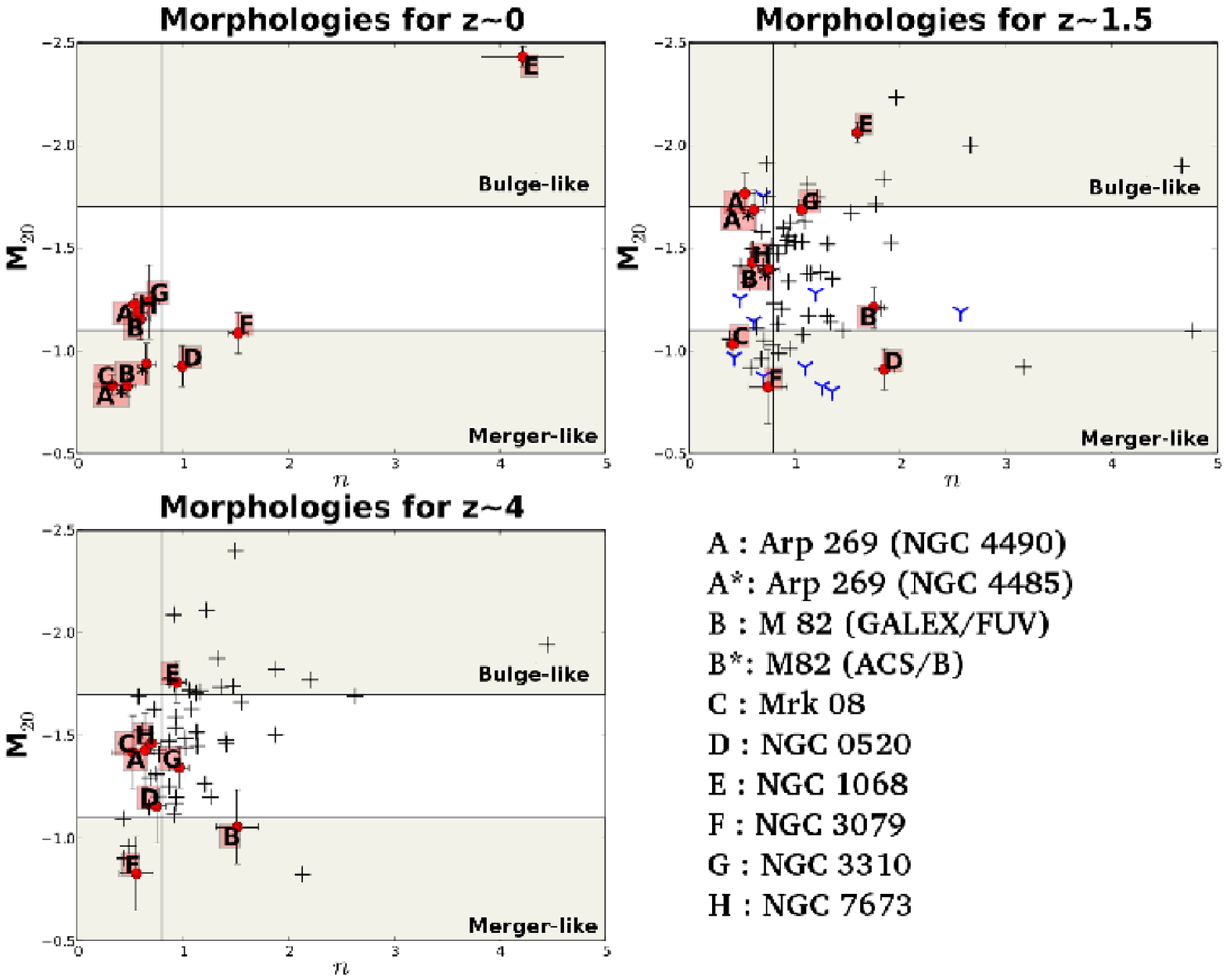}
\caption{We plot $M_{20}$ and $n$, S\'ersic index, to determine high redshift star-forming galaxy analogs (see \S \ref{analogs}). The symbols are the same as for Figure \ref{morphplot}. The shaded regions $0<$M$_{20}<-1.1$ and $-1.7<$M$_{20}<-2.5$ distinguish between merger- and bulge-dominated galaxies. The vertical line marks $n=0.8$, which helps distinguish whether the object, if to the left of the line, has a merger profile. The error bars for $n$ values are based on \cite{rav06}. We note that 7 of the 10 galaxies are below $n=0.8$ in the FUV and move to various locations in the plot as the systems are redshifted. NGC 1068 (E) is especially noteworthy, because it moves from an exponential profile to right along the border of the $n=0.8$/M$_{20}=-1.7$ line as it is redshifted. We determine that Mrk 08, NGC 7673 and NGC 3079 are high redshift star-forming galaxy analogs based on their morphologies presented in this figure and in Fig. \ref{morphplot}.
}\label{galmorpha}
\end{figure*}
}

\def\morphb{
\begin{figure*}[p]
\centering
\epsscale{2.2}
\plotone{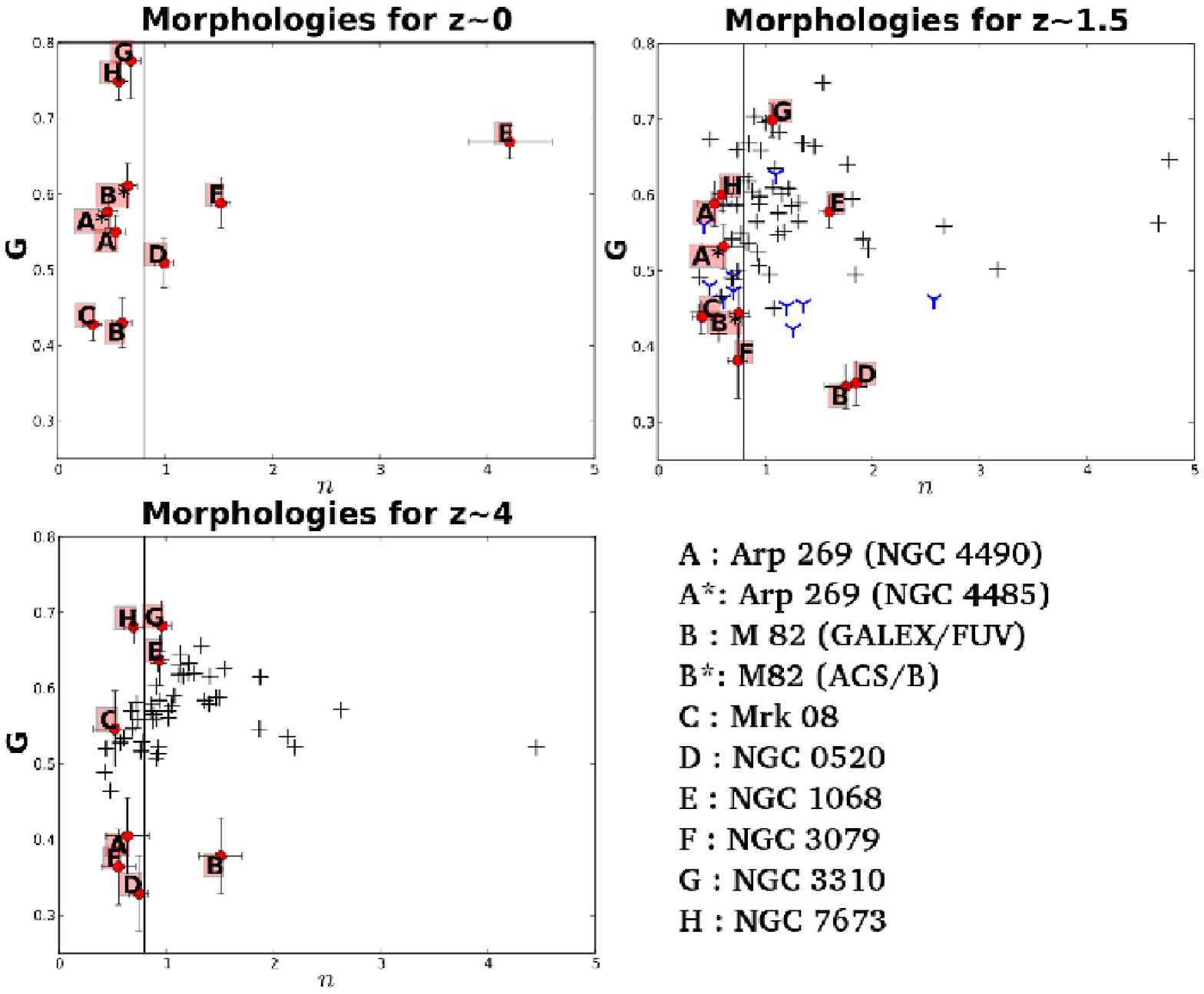}
\caption{We plot G and $n$, S\'ersic index, to determine high redshift star-forming galaxy analogs (see \S \ref{analogs}). The symbols are the same as for Figure \ref{morphplot}. The vertical line marks $n=0.8$, which helps distinguish whether the object, if to the left of the line, has a merger profile.
}\label{galmorphb}
\end{figure*}
}

\def\gentable{
\begin{deluxetable}{lcccccc}
\tabletypesize{\footnotesize}
\tablecaption{Basic Data for Nearby Galaxy Sample\tablenotemark{a}}
\tablenum{1}
\tablewidth{0pt}
\tablehead{
\colhead{Name} & \colhead{Alt. Name} & \colhead{$v$}\tablenotemark{b} & \colhead{Magnitude}\tablenotemark{c} & \colhead{Morphology} & \colhead{Diameter}\tablenotemark{d} & \colhead{Luminosity}\tablenotemark{e}\\
\colhead{} & \colhead{} & \colhead{$km s^{-1}$} & \colhead{B$_{\rm{Tot}}$} & \colhead{} & \colhead{$\arcmin$} 
& \colhead{$L_{\sun}$}
}
\startdata
Arp 269 (NGC 4490) & UGC 07651 & 565$\pm$3 & 10.22$\pm$.06 & SB(s)d pec & 6.3 & $\sbond$
\\
Arp 269 (NGC 4485) & UGC 07648 & 493$\pm$7 & 12.32$\pm$.05 & IB(s)m pec & 2.3 & $\sbond$ \\
M 82 & NGC 3034 & 203$\pm$4 & 9.30$\pm$.09 & I0:Sbrst & 11.2 & $6.97\times10^{7}$ \\
Mrk 08 & UGC 03852 & 3604$\pm$8 & 14.0$\pm$.2 & I? & 0.8 & $9.42\times10^{8}$ \\
NGC 0520 & UGC 00966 & 2281$\pm$3 & 12.24$\pm$.13 & pec; Sbrst & 4.5 & $2.40\times10^{8}
$\\
NGC 1068 & M 77 & 1137$\pm$3 & 9.61$\pm$.1 & (R)SA(rs)b; Sy1/y2 & 7.1 & $1.53\times10^{9}$
\\
NGC 3079 & UGC 05387 & 1116$\pm$1 & 11.54$\pm$.14 & SB(s)c; Sy2 & 7.9 & $4.02\times10^{8}
$ \\
NGC 3310 & UGC 05786 & 993$\pm$3 & 11.15$\pm$.1 & SAB(r)bc pec & 3.1 & $\sbond$\\
NGC 7673 & UGC 12607 & 3408$\pm$1 & 13.17$\pm$.13 & (R')SAc? pec & 1.3 & $2.64\times10^{9}
$ \\
\enddata
\tablenotetext{a}{NASA/IPAC Extragalactic Database (NED) and RC3 catalogue \citep{vau91}}
\tablenotetext{b}{Radial velocity}
\tablenotetext{c}{Total B Vega magnitude}
\tablenotetext{d}{Major axis}
\tablenotetext{e}{The GALEX Ultraviolet Atlas of Nearby Galaxies \citep{gil07}}
\end{deluxetable}

}

\def\fuvtable{
\begin{deluxetable}{lccccccccccccccl}
\rotate
\tabletypesize{\scriptsize}
\tablecaption{Morphologies for Artificially Redshifted Sample}
\tablenum{2}
\tablewidth{0pt}
\tablehead{
\colhead{Object} & \colhead{G$^{z0}$} & \colhead{G$^{z1.5}$} & \colhead{G$^{z4}$} & \colhead{$M^{z0}_{20}$} & \colhead{$M^{z1.5}_{20}$} & \colhead{$M^{z4}_{20}$} & \colhead{$r^{z0}_{p}$} & \colhead{$r^{z1.5}_{p}$} & \colhead{$r^{z4}_{p}$} & \colhead{$n^{z0}$} & \colhead{$n^{z1.5}$} & \colhead{$n^{z4}$} \\ 
\colhead{} & \colhead{} & \colhead{} & \colhead{} & \colhead{} & \colhead{} & \colhead{} & \colhead{$\arcsec$} & \colhead{$\arcsec$} & \colhead{$\arcsec$} & \colhead{} & \colhead{} & \colhead{} 
} 
\startdata
Arp 269 (NGC 4490) & 0.55$\pm$0.02 & 0.59$\pm$0.03 & 0.41$\pm$0.05 & -1.23$\pm$0.05 & -1.77$\pm$0.10 & -1.43$\pm$0.18 & 128.0 & 0.71 & 0.30 & 0.54$\pm0.09$ & 0.53$\pm0.16$ & 0.64$\pm0.2$   \\
Arp 269 (NGC 4485) & 0.58$\pm$0.02 & 0.53$\pm$0.03 & $\sbond$ & -0.83$\pm$0.05 & -1.69$\pm$0.10 & $\sbond$ & 47.4 & 0.50 & $\sbond$ & 0.47$\pm0.09$ & 0.61$\pm0.20$ & $\sbond$ \\
M82 (FUV) & 0.43$\pm$0.03 & 0.35$\pm$0.03 & 0.38$\pm$0.05 & -1.16$\pm$0.10 & -1.21$\pm$0.10 & -1.05$\pm$0.18 & 241.2 & 1.07 & 0.24 & 0.60$\pm0.09$ & 1.75$\pm0.20$ & 1.51$\pm0.20$ \\
M82 (ACS B$_{435}$) & 0.61$\pm$0.03 & 0.44$\pm$0.03 & $\sbond$ & -0.94$\pm$0.10 & -1.40$\pm$0.10 & $\sbond$ & 23.9 & 1.11 & $\sbond$ & 0.65$\pm0.09$ & 0.76$\pm0.09$ & $\sbond$  \\
Mrk 08 & 0.43$\pm$0.02 & 0.44$\pm$0.02 & 0.55$\pm$0.05 & -0.83$\pm$0.05 & -1.03$\pm$0.05 & -1.42$\pm$0.18 & 21.8 & 0.66 & 0.40 & 0.33$\pm0.09$ & 0.41$\pm0.09$ & 0.53$\pm0.20$  \\
NGC 0520 & 0.51$\pm$0.03 & 0.35$\pm$0.03 & 0.33$\pm$0.05 & -0.92$\pm$0.10 & -0.91$\pm$0.10 & -1.16$\pm$0.18 & 77.4 & 1.13 & 0.46 & 0.99$\pm0.09$ & 1.85$\pm0.09$ & 0.75$\pm0.09$  \\
NGC 1068 & 0.67$\pm$0.02 & 0.58$\pm$0.02 & 0.64$\pm$0.03 & -2.43$\pm$0.05 & -2.06$\pm$0.05 & -1.76$\pm$0.10 & 26.7 & 0.56 & 0.24 & 4.22$\pm0.39$ & 1.60$\pm0.09$ & 0.94$\pm0.09$  \\
NGC 3079 & 0.59$\pm$0.03 & 0.38$\pm$0.05 & 0.36$\pm$0.05 & -1.09$\pm$0.10 & -0.82$\pm$0.18 & -0.83$\pm$0.18 & 244.3 & 2.44 & 1.12 & 1.52$\pm0.09$ & 0.75$\pm0.09$ & 0.56$\pm0.16$ \\
NGC 3310 & 0.78$\pm$0.05 & 0.70$\pm$0.02 & 0.68$\pm$0.03 & -1.24$\pm$0.18 & -1.69$\pm$0.05 & -1.34$\pm$0.10 & 33.6 & 0.42 & 0.25 & 0.68$\pm0.09$ & 1.07$\pm0.09$ & 0.96$\pm0.09$\\
NGC 7673 & 0.75$\pm$0.02 & 0.60$\pm$0.02 & 0.68$\pm$0.02 & -1.18$\pm$0.07 & -1.43$\pm$0.05 & -1.46$\pm$0.05 & 17.6 & 0.42 & 0.22 & 0.57$\pm0.09$ & 0.60$\pm0.09$ & 0.70$\pm0.09$\\
\enddata
\end{deluxetable}

}
\def\mybib{

\clearpage
}

\def\Bbnd{B$_{435}$}
\def\Vbnd{V$_{555}$}
\def\ibnd{$i_{775}$}
\def\zbnd{$z_{850}$}

\shorttitle{Comparing Local Starbursts}
\shortauthors{Petty et al.}
\slugcomment{Submitted to AJ, revised after Referee comments}

\begin{document}

\title{Structures of Local Galaxies Compared to High Redshift Star-forming Galaxies}

\author{Sara M. Petty,\altaffilmark{1,2} Du\'ilia F. de Mello,\altaffilmark{1,2,3} John S. Gallagher III,\altaffilmark{4} Jonathan P. Gardner,\altaffilmark{2} Jennifer M. Lotz,\altaffilmark{5} C. Matt Mountain,\altaffilmark{6} Linda J. Smith\altaffilmark{6,7,8}}

\altaffiltext{1}{Department of Physics, The Catholic University of America, Washington, DC 20064; sara.m.petty@nasa.gov}
\altaffiltext{2}{Observational Cosmology Laboratory, Code 665, Goddard Space Flight Center, Greenbelt, MD 20771}
\altaffiltext{3}{Johns Hopkins University, Baltimore, MD 21218}
\altaffiltext{4}{Department of Astronomy, University of Wisconsin-Madison, 475 North Charter Street, Madison, WI 53706}
\altaffiltext{5}{National Optical Astronomical Observatory, 950 N. Cherry Ave., Tucson, AZ 85719}
\altaffiltext{6}{Space Telescope Science Institute, 3700 San Martin Dr., Baltimore, MD 21218}
\altaffiltext{7}{ESA Space Telescope Operations Division, MD 21218, USA}
\altaffiltext{8}{Department of Physics and Astronomy, University College London, Gower Street, London WC1E 6BT, UK}

%%%%%%%%%%%%%%%%%%%%%%%%%%%%%%%%ABSTRACT%%%%%%%%%%%%%%%%%%%%%%%%%%%%%%%%%
\begin{abstract}
The rest-frame far-ultraviolet (FUV) morphologies of 8 nearby interacting and starburst galaxies (Arp 269, M 82, Mrk 8, NGC 520, NGC 1068, NGC 3079, NGC 3310, \& NGC 7673) are compared with 54 galaxies at z$\sim1.5$ and 46 galaxies at z$\sim4$ observed in the Great Observatories Origins Deep Survey (GOODS) taken with the Advanced Camera for Surveys onboard the Hubble Space Telescope. The nearby sample is artificially redshifted to z$\sim1.5$ and 4 by applying luminosity and size scaling. We compare the simulated galaxy morphologies to real z$\sim1.5$ and 4 UV-bright galaxy morphologies. We calculate the Gini coefficient (G), the second-order moment of the brightest 20\% of the galaxy's flux ($M_{20}$), and the S\'ersic index ($n$). We explore the use of nonparametric methods with 2D profile fitting and find the combination of $M_{20}$ with $n$ an efficient method to classify galaxies as having merger, exponential disk, or bulge-like morphologies. When classified according to G and $M_{20}$ 20/30\% of real/simulated galaxies at z $\sim1.5$ and 37/12\% at z$\sim4$ have bulge-like morphologies. The rest have merger-like or intermediate distributions. Alternatively, when classified according to the S\'ersic index, 70\% of the z$\sim1.5$ and z$\sim4$ real galaxies are exponential disks or bulge-like with $n>0.8$, and $\sim30\%$ of the real galaxies are classified as mergers. The artificially redshifted galaxies have $n$ values with $\sim35\%$ bulge or exponential at z$\sim1.5$ and 4. Therefore, $\sim20-30\%$ of Lyman-break galaxies (LBGs) have structures similar to local starburst mergers, and may be driven by similar processes. We assume merger-like or clumpy star-forming galaxies in the GOODS field have morphological structure with values $n<0.8$ and $M_{20}>-1.7$. We conclude that Mrk 8, NGC 3079, and NGC 7673 have structures similar to those of merger-like and clumpy star-forming galaxies observed at z$\sim1.5$ and $4$.
\end{abstract}


% % % % % % % % % % % % % % % % % % %Start Main Body% % % % % % % % % % % % % % % % % % %
\section{Introduction}

Deep images of the universe have shown that many of the progenitors of present-day galaxies are experiencing very active star formation and undergoing violent gravitational interactions. However, the method by which interactions in youthful galaxies drive their baryonic structures is still an open discussion. Moreover, it is still not clear in which respects the star formation processes at high-redshift, differ from current galaxies. One way of trying to visualize how these processes influence our view of the distant universe is by artificially placing local interacting and starburst galaxies at high redshift and comparing their properties with observed high redshift objects.

Pinpointing an epoch for the formation of the Hubble sequence is difficult, even though a few observational studies suggest that it occurred in the redshift range $1<$ z $<2$ \citep{con04, pap05}. The Hubble sequence classifies galaxies mostly by bulge to disk ratios, but this fails at higher redshifts. At high redshift, the color dispersion, size and luminosity of  galaxies from z$\sim3$ to $\sim1$ are important gauges of galaxy evolutionary processes. \cite{pap05} find that the color dispersion is higher for galaxies at z$\sim1$, than for galaxies at z$\sim3$, implying a lack of older stellar populations in the higher redshift galaxies. However, other studies have shown a significant population of extremely red objects (EROs), or distant red galaxies (DRGs) at z$> 1.5$ \citep[e.g.][]{lab03,fra03,gla04,dad05,wik08}. It is unknown what mechanisms turn off and continue to prevent star formation in massive galaxies at early times.

Relating the star formation history to the galaxy size and distribution of light is a primary goal of evolutionary studies. The reduction in the effective radii and the luminosity increase of galaxies at z $\ga1.5$ have been well studied and confirmed by many authors \citep[e.g.][]{fer04,arn05,tof07,aki08,fra08}. In particular, \cite{fra08} show that size and luminosity evolution is less a function of stellar mass and more a function of surface brightness distribution. \cite{tof07} show a correlation between surface mass densities and star-forming galaxies with redshift. By using surface mass densities from derived stellar masses of their sample, the z$>$ 2.5 star-forming galaxies on average have a larger surface mass density (factor of $\sim$ 6) than local star-forming galaxies of similar masses. \cite{pap05} find that the mean galaxy size, of UV-bright galaxies, increases by $40\%$ from z$\sim2.3$ to $\sim1$ and that characteristic sizes have not changed since z$\sim1$.  A consideration of size and luminosity with increasing redshift is imperative when comparing local galaxies with high redshift galaxies.

Star-forming galaxies, such as Lyman-break galaxies (LBGs), are the focus of numerous studies particularly for the purpose of determining the star formation history of galaxies with redshift \citep[see][]{mad96}. LBGs are z $\ga2.5$ U- and B-dropout galaxies selected by the colors of their FUV-rest-frame spectral energy distribution around the 912$\rm{\AA}$ Lyman continuum discontinuity \citep[see][]{ste96}. The lack of star-forming galaxies in the redshift desert ($1.4\la$ z $\la2.5$) recently presented a great puzzle. \cite{ste04} and \cite{sav04} show this is simply due to lack of observations. \citeauthor{ste04} color-selected lower redshift Lyman-break galaxies (LBGs) and followed-up with spectroscopic observations, showing that LBGs are present at lower redshifts than originally thought.

Comparison spectra of LBGs with local starburst galaxies have shown remarkable similarities \citep{pet00,mel00}. For example, the well-studied MS 1512-cB58 (z$\sim3$) shows absorption line features typical of nearby starburst galaxies, which is reproduced with stellar synthesis models. However, the UV luminosities and star-formation rates are significantly larger for LBGs than for local starburst galaxies \citep{gia02}. This raises questions about how the morphologies of high redshift star-forming galaxies resemble local starburst galaxies, and how the mechanisms driving star formation in high redshift galaxies such as LBGs are unique to galaxy formation in the early universe.

In order to answer these questions, many different approaches to connect high redshift star-forming galaxies with nearby galaxies have been undertaken. \cite{hoo07} used stellar mass ratios, surface brightness and luminosity plots to find ultraviolet luminous galaxies (UVLGs; L $> 2 \times 10^{10}$L$_{\odot}$) that fit the typical LBG surface brightness profile. According to \cite{hec05}, such UVLGs are very compact with large surface brightnesses (L $> 10^{8}$L$_{\odot}$ kpc$^{-2}$), but this does not tell us whether these systems have multiple clumps, or single bulges, nor do they determine the type of surface brightness profile. In a thorough investigation of this, \cite{over08} studied the bulge, disk, and clumpy nature of 8 UVLGs. They also artificially redshifted them and determined that they fit the morphology for high redshift starburst galaxies and are good LBG-analogs. The presence of starburst clumps in LBGs were particularly discussed as a sign of patchy compact star-forming regions due to merging. The artificial redshifting tests in the \citeauthor{over08} study show that luminous star clusters within a galaxy at high redshift can dominate the luminosity. 

One of the first attempts at artificially redshifting rest-UV images of nearby starburst galaxies was done by \cite{hib97} \citep[see][for the first study]{wee85}. The authors use B-, and V-band images to simulate the original Hubble Deep Field \citep[HDF][]{wil96} at z$\sim0.5-2.5$. They particularly look at peculiar galaxies and find that the tidal features from these disturbed galaxies are still viewable at high redshift. They do not apply evolutionary effects, such as luminosity or size evolution to their sample. They warn of the biases of measuring morphologies for z$>1.5$ systems, since only the regions that have been least extinguished are detected. More recently, the Advanced Camera for Surveys (ACS), covering a larger area in the Great Observatories Origins Deep Survey (GOODS), provided better statistics in the field of high redshift morphology.

\cite{lotz06} use the Gini coefficient (G), the second-order moment of the brightest 20\% of the galaxy's flux ($M_{20}$), and concentration (C) morphology analysis to classify rest-frame FUV emission from galaxies at z$\sim1.5$ and 4 with GOODS and Hubble Ultra Deep Field (HUDF) images. The authors created a noise-free de Vaucouleurs bulge and exponential disk to use as benchmarks with the z$\sim1.5$ and 4 samples. They find that $\sim30\%$ of the LBGs are bulge-like in their morphologies, $\sim50\%$ have values closer to merger-like or clumpy star forming regions, and 10-25\% of the LBG sample have morphologies consistent with ongoing major mergers. Their results are roughly in agreement with hierarchical model predictions for merger rates at high redshifts in \cite{som01}, since the z$\sim1.5$ sample has more extended star-forming disks than the z$\sim4$ sample. Quantitative comparisons such as the kinematic, non-parametric and 2D profiles of high redshift objects are difficult to obtain, because of resolution and low S/N levels compared with the local counterparts. Artificially redshifting local galaxies is a way to place local galaxies at a similar resolution and conduct morphological studies.

\cite{rav04,rav06} established surface brightness profiles, using the S\'ersic index ($n$), and ellipticities of high redshift galaxies. In \cite{rav06}, a multiwavelength study was conducted using GOODS images and identified $\sim4700$ LBGs and $292$ starburst galaxies at z$\sim1.2$. Exponential profiles comprised about 40\% of the LBGs, while $\sim30\%$ have steep, $r^{1/4}$-like, profiles and $\sim30\%$ have disturbed morphologies. Recently, \cite{raw09} performed a similar study to compare the rest-frame UV and rest-frame optical single component S\'ersic fits. The authors deduced that the S\'ersic index is smaller in the rest-frame UV than in the rest-frame optical, especially for the clumpy or merging galaxies in their sample.

In this study, we present the morphologies of 8 starburst galaxies observed in FUV ($\lambda \sim 1500\rm{\,\AA}$), using G, $M_{20}$, and the S\'ersic index ($n$) and we compare the values of these objects with GOODS z$\sim1.5$ and 4 FUV-rest-frame galaxies. We attempt to observe how distant galaxies might look if they had the structures of appropriately scaled versions of nearby galaxies. As objects are artificially redshifted, we classify them based on the clumpiness of the disk and bulge structures. A key point in our study is that we include a combination of nonparametric and 2D S\'ersic profile fitting to base our conclusions. Other studies have focused on one type of method to derive standards by which to classify galaxies \citep[eg.][]{lotz06,rav06,over08}. We combine both methods for a more complete quantitative analysis. In \S \ref{sample} and \S \ref{artred}, we describe our sample and approach to simulating high redshifts. In \S \ref{techniques}, we explain in detail the analysis techniques used to quantitatively classify the morphologies, and \S \ref{results} provides the results of our study. Finally, we conclude with our key discoveries in \S \ref{summary}. We have adopted the cosmological constants $\rm{H_{0}}=70 \rm{\,km \,s^{-1}\, Mpc^{-1}}, \Omega_{\rm{m}}=0.3,$ and $\Omega_{\rm{\Lambda}}=0.7$ throughout this paper.

%  % % % % % % % % % % % % % % % % % % % % % % % %DATA & METHODS%%%%%%%%%%%%%%%%%%%%%%%%%%%%%%%%%%%%%%%
\section{Galaxy Sample \label{sample}}

\subsection{Nearby Starbursts}
Our sample consists of the following: three interacting galaxies with well extended HI tidal tails from Hibbard et al. (2001) --Arp 269, M 82, NGC 3079; three starburst galaxies with varied morphologies --Mrk 8, NGC 3310, NGC 7673; and two galaxies from \cite{lotz06} --NGC 520, NGC 1068 for comparison with \citeauthor{lotz06} results. We selected this sample for the variety of morphologies and characteristics that it represents and the availability of appropriate FUV data. Starburst galaxies are important, since LBGs are starburst galaxies by selection. Interacting galaxies tend to have more disturbed morphologies and might be more representative of the clumpy star-forming galaxy populations at high redshift. As this is a pilot study to explore techniques, we do not intend for our local galaxy sample to be statistically representative, but rather to cover common starburst galaxy morphologies.

We use the GALEX/FUV ($\lambda_{\rm{eff}} = 1516\, \rm{\AA}$) sky-subtracted images provided by Multimission Archive at STScI (MAST) observed by the GALEX Nearby Galaxy Survey. Figures \ref{imgsa} and \ref{imgsb} show the optical images in column (a) (Sloan Digital Sky Survey \citep{yor00}, or Digitized Sky Survey as labeled) and GALEX/FUV in column (b). We artificially redshifted GALEX/FUV images for all galaxies. In the case of M 82, we also included ACS F435W (\Bbnd) artificially redshifted to z$\sim1.5$, which corresponds with observations in the ACS F850L (\zbnd). The images for simulating the ACS F435W, F555W (\Vbnd), F775W (\ibnd) and F850L backgrounds, were taken from empty sky regions in the GOODS field. We describe the artificial redshift process in \S \ref{artred}. We provide general information for our objects in Table 1. 

\subsection{GOODS Comparison Sample at z$\sim 1.5$ to 4}
For our actual high redshift galaxy sample, we selected starburst galaxies from two previous studies. We chose the 10 brightest $0.8 \la z \la 1.8$ galaxies (out of 94) from the \cite{voy} U-band catalog in the Ultra Deep Field, giving rest-frame FUV at this redshift range. We use these 10 galaxies to compare with the artificially redshifted sample. We created GOODS/ACS \Bbnd-band cutouts from MAST and analyzed their morphologies as described in \S \ref{techniques}.

The other comparison sample of galaxies at z$\sim1.5$ and 4 comes from \cite{lotz06}. All of the GOODS sample were used since they fit our criteria of galaxies bright in the rest-frame FUV at z $\sim 1.5$ to 4. An added benefit has been the ability to test our method of morphology measurements with \cite{lotz06} results. We included 54 starburst galaxies from the GOODS North and South fields in the redshift range 1.2$<$z$<$1.8. The z$\sim4$ LBGs (46) are GOODS South B-dropouts. We acquired the GOODS/ACS \Bbnd-, \Vbnd-, \ibnd- and \zbnd-band images of these objects from MAST.

\section{Artificial Redshift Simulations\label{artred}}

We artificially redshift GALEX/FUV images of the local starburst galaxies to z$\sim1.5$ and $4$ as if they were observed in the GOODS/ACS \Bbnd- and \Vbnd+\ibnd-bands. A section of background sky in the GOODS field for each of the bands was cut and added to the artificially redshifted object. To redshift the GALEX/FUV images, we used the size evolution relation discussed in \cite{fer04}, and luminosity function for FUV galaxies measured by \cite{arn05} in order to compare with LBGs. We describe the method in this section.

For size evolution, z$\sim4$ objects have a half-light radii reduction of $\sim0.4$ ($\rm{r_{highz}} = 0.4 \, \rm{r_{lowz}}$). To accomplish this, we determined the rebinning factor, $N_{\rm{highz}}$, which has the form
\begin{equation}
 N_{\rm{highz}} = N_{\rm{lowz}}\ast\sigma\ast n
\end{equation}
where $N_{\rm{lowz}}$ is the number of pixels in one dimension in the original low redshift image, $n$ is the size evolution factor (we use 0.4 for z $\sim4$), and $\sigma$ is the scale ratio determined by
\begin{equation}
 \sigma = \frac{\theta_{\rm{lowz}}}{\theta_{\rm{highz}}}\frac{\phi_{\rm{lowz}}}{\phi_{\rm{highz}}}.
\end{equation}
$\theta_{\rm{lowz}}$ and $\theta_{\rm{highz}}$ are the angular sizes for low and high redshift, and $\phi_{\rm{lowz}}$ and $\phi_{\rm{highz}}$ are the pixel scales of the images \citep[see][]{gia96,lotz06}. The angular size depends on redshift as defined in Equation 2 in \cite{gia96} as
\begin{equation}
 \theta_{\rm{z}} = d \frac{\rm{(1+z)}^2}{\rm{L_{z}}}.
\end{equation}
L$_{\rm z}$ and $d$ are the luminosity distance and physical size of the galaxy. 

We also account for the pixel scaling similar to the method described in Equations 6 and 7 in \cite{gia96}. Where a conversion of count rates is required to account for the apparent magnitude conversion due to L$_{\rm z}$ and angular distance ($D$). Where $D$ is defined as
\begin{equation}
 D = \frac{\rm{(1+z)}^2}{\rm{L_{z}}}.
\end{equation}
The dimming is applied by multiplying the low redshift pixel counts ($p_{\rm{lowz}}$) by a factor of L$^2_{\rm{lowz}}$/L$^2_{\rm{highz}}$ and with luminosity evolution the high redshift pixel counts ($p_{\rm{highz}}$) have the form
\begin{equation}
 p_{\rm{highz}} \propto p_{\rm{lowz}}\, \rm{X_{boost}} \,\frac{\rm{L_{\rm{lowz}}^2}}{L_{\rm{highz}}^2},
\end{equation}
where $\rm{X_{boost}}$ is the luminosity evolution factor described below. 

The luminosity evolution for FUV-rest-frame galaxies with z$\sim1.5$ to $4$ gives a change in magnitude $\Delta \rm{M}\sim-3$, since the star formation rates of earlier galaxies is much higher than in the present day. To apply this to the image to be redshifted, we needed to find the value to multiply, or boost the original low redshift image. This was done by using the counts per second (CPS) to magnitude (AB) conversion for GALEX FUV:
\begin{equation}
 \rm{M_{AB}} = -2.5 log10(CPS) + 18.82
\end{equation}
 For $\Delta \rm{M}\sim-3$ the boosting factor becomes
\begin{equation}
 \rm{X_{\rm{boost}}} = 10^{\frac{3}{2.5}} \sim 15.
\end{equation}
We multiplied the low redshift pixels by this amount for both the z$\sim1.5$ and $4$ images as described above in Equation 5. For artificially redshifting M 82 using ACS \Bbnd \hspace{1pt} to \zbnd \hspace{1pt} (simulating z$\sim1.5$) we boosted the magnitudes by $\Delta M\sim-1$. This accounted for the luminosity evolution in the general population at high redshift in the optical rest-frame \citep[see][]{rav04,bar05,cam07}.

To artificially redshift nearby galaxies to higher redshifts, the original images are sky subtracted, then rebinned by the size and scale ratios discussed above, and boosted by the appropriate factor. The resulting image is convolved with a PSF created from the appropriate GOODS/ACS band and then added to a background sky image created from an empty region in the GOODS \Bbnd-, or \Vbnd+\ibnd-bands \citep[see][]{wee85,gia96,lotz06,over08}. 

We use the size and luminosity assumptions for two reasons. One is that the local galaxies would be too faint at z$\sim 1.5$ and 4 to measure their morphologies, so the magnitudes need to be boosted. The second reason is that the compact sizes of high redshift galaxies may bias their morphologies. Including size evolution is imperative for a fair comparison. Additionally, our sample, unlike the \cite{over08} and \cite{hoo07} sample of UVLGs, does not have UV luminosities and sizes comparable to LBGs. We did not choose the sample based on high redshift star-forming galaxy analogs. Rather, we chose examples of nearby starbursts covering a range in structure. These galaxies are scaled to become simulations of high redshift star-forming galaxies in terms of size and surface brightness while keeping the same internal structures. We are asking what we know about the structure of high redshift galaxies and showing that it may be less than currently understood.

\section{Morphological Analysis \label{techniques}}
We used SExtractor \citep{ber96} for object detection and to create object segmentation maps with a detection threshold set to $0.6 \sigma$. The object centers, and Petrosian radius were then used to determine the Gini coefficient, $M_{20}$, and to create GALFIT models. We defined $1.5r_{\rm{p}}$ ($r_{\rm{p}}$ is the Petrosian radius at which the ratio of the surface brightness to the mean interior surface brightness is $\eta = 0.2$) as the object radius, following the method by \cite{lotz06} \citep[see][for a discussion regarding other methods]{abr07,law07}. For a benchmark, \cite{lotz06} created a noise-free de Vaucouleurs bulge and exponential disk and find: G $= 0.6$ and $M_{20}=-2.47$ for the bulge; and G $=0.473$ and $M_{20}=-1.80$ for the exponential disk.

\subsection{The Gini Coefficient}

One way to quantitatively determine the stellar structures of galaxies is through nonparametric analysis, such as G and $M_{20}$ \citep[e.g.][]{abr03,lotz04}, where the primary goal is to identify merging and interacting galaxies. The Gini coefficient is correlated with concentration and surface brightness. The Gini coefficient is defined as 
\begin{equation}
\rm{G} = \frac{1}{\displaystyle |\bar{X}|l(l-1)}\sum_{i=1}^{l}(2i-l-1)|X_{i}|.
\end{equation}
The sum is from $i = 1$ to $l$, where $\bar{X}, X_{i}$, and $l$ are the mean flux, the rank ordered pixel flux values, and total number of pixels in the object map, respectively. Note that $X_{i}$ is sorted in increasing order. For $\rm{G}=1$, all of the light resides in one pixel, and for $\rm{G}=0$ all of the light is evenly distributed  between the pixels. According to \cite{lotz04}, typical Gini coefficient values in the FUV are: $\sim0.35$ to $0.55$ for merger- or disk/transition; $\sim0.55$ to $0.65$ for bulge-dominated morphological types. 
 
\subsection{$M_{20}$}

$M_{20}$ is a logarithmic ratio giving the normalized second-order moment of the brightest 20\% of the
total flux of the galaxy. Therefore, the more negative value for $M_{20}$, the more centrally concentrated the object. $M_{20}$ is defined by
\begin{equation}
M_{20} \equiv {\rm log10}\left(\frac{\sum_i M_i}{M_{\rm tot}}\right) {\rm with } \sum_i X_i <  0.2 X_{\rm tot},
\end{equation}
where $M_{i}$ is given by
\begin{equation}
M_{i} = X_{i}[(x_{i}-x_{c})^{2}+(y_{i}-y_{c})^{2}].
\end{equation}
$M_{\rm{tot}}$ is the sum from $i = 1$ to $l$, and $x_{c}$, and $y_{c}$ are the object centers in the $x$ and $y$ position. $M_{i}$ is summed over the brightest pixels in 20\% of the total flux ($X_{\rm tot}$). Typical $M_{20}$ values are approximately $-0.8$ to $-1.1$ for merger-like, and $-1.7$ to $-2.2$ for bulge-dominated morphological types. The merger-like values come from detecting multiple nuclei.

\subsection{S\'ersic 2D modeling}
Another morphological analysis method is through 2D modeling, such as the S\'ersic index, $n$, which fits a range of distributions from exponential disks ($n = 1$) to $r^{1/4}$ spheroids ($n = 4$). We used GALFIT software \citep{peng02} to model 2D profiles with the S\'ersic index. When $n$ is large, the inner profile is steep and the outer profile is extended. When $n$ is small, the inner profile is shallow and has a steep cut-off at a large radius. Monte Carlo simulations by \cite{rav06} give $<n>=3.83$ for spheroids, and $<n>=1.1$ for disks, which allows for the broad classification scheme of $n>2.5$ for spheroids, and $n<2.5$ for disks. This scheme may be further divided where typical values for $n$ are: $n < 0.8$ for mergers; $0.8 < n < 2.5$ for exponential profile systems; $n > 2.5$ for bulge systems \citep{rav06}.

%%%%%%%%%%%%%%%%%%%%%%%%%%%%RESULTS%%%%%%%%%%%%%%%%%%%%%%%%%%%%
\section{Results \label{results}}
Figures \ref{imgsa}, \ref{imgsb}, and \ref{galfita} show the artificially redshifted sample and the resulting GALFIT model images. In Figures \ref{imgsa} and \ref{imgsb}, the four columns divide the data by (from left to right): (a) DSS or SDSS color composite images; (b) GALEX/FUV; (c) simulation of ACS-\Bbnd \hspace*{1pt} z$\sim1.5$; (d)  simulation of ACS-\Vbnd+\ibnd \hspace*{1pt} z$\sim4$. M 82 was artificially redshifted with two different wavelength images, to emphasize the change in morphology from UV to optical. We artificially redshifted the ACS-\Bbnd \hspace{1pt} image into the \zbnd-band as shown in columns (b) and (c). Note the scale change between the z$\sim1.5$ and z$\sim4$ images. We have drawn 1$\arcsec$ rulers on the z$\sim1.5$ and z$\sim4$ images and 1$\arcmin$ rulers on the z$\sim0$ images to highlight this difference.

The object name, morphological parameters G, $M_{20}$, $n$, and $r_{\rm{p}}$ are displayed in Table 2 for each artificially redshifted object. Every parameter has three listings for the separate redshifts, described by the superscripted z0, z1.5, and z4. The measurement of uncertainties for G and $M_{20}$ are inferred from the average S/N per pixel within the Petrosian radius of the galaxy images and Figure 1 of \cite{lotz06}. The authors found that the typical dispersion in the difference between the G/$M_{20}$ measurements in very deep Ultra Deep Field and shallower GOODS images of the same object was a function of the object's average signal-to-noise per pixel in the GOODS observations. \cite{lotz04} explored the effect of morphological type on G/$M_{20}$ measurement error, and found that morphological type had a negligible effect when average signal-to-noise per pixel was greater than 2. Uncertainties for $n$ are based on \cite{rav06}, where the error is determined by effective radius, and magnitude of a disk ($n<2.5$) or spheroid ($n>2.5$).

We discuss each nearby starburst object separately in \S 5.1, and provide the quantitative results for each redshift (z $\sim0, \sim1.5, \& \sim4$) in \S 5.5. We also explain how we determined the high redshift star-forming galaxy analogs in \S \ref{analogs}.

\subsection{Artificially Redshifted Starburst Galaxies}

The properties of the galaxies in Figures \ref{imgsa} and \ref{imgsb} are qualitatively described below.

\begin{flushleft}
\textbf{Arp 269}: This is an interacting minor merger which we selected from \cite{hib01}. It consists of the interaction between NGC 4485 and NGC 4490 with an extended HI tail where tidal effects can be seen. This Arp galaxy has mostly been studied as a part of larger samples in the radio through X-ray wavelengths to compare properties of interacting galaxies with single galaxies \citep[e.g.,][]{you86,bok99,cle02,cas04}. The GALEX/FUV image clearly shows the two separate galaxies. As the system is redshifted and scaled to z$\sim4$ SExtractor does not deblend the two galaxies, and it is detected as a single object.\\
\textbf{M 82}: This strong starburst galaxy is a well studied member of the M 81 interacting triplet (M 81, M 82, and NGC 3077; also selected from \cite{hib01}). The HI tidal tail is well extended from M82. It has very different characteristics in the FUV compared with the optical. The optical image looks like a typical edge-on galaxy. The FUV has a clover pattern, due to the hot wind cones coming from the nucleus, making it difficult to measure the surface brightness of the disk. The FUV emission is due to reflection in the dusty wind. The cones are visible in H$\alpha$ and are barely resolved in the artificially redshifted images, because of their intrinsically low surface brightness. The extended gas disappears within the noise and leaves a small, peculiar shaped object. The optical continues to look like an edge-on in the \Bbnd- to \zbnd-band redshift image, while the rest-frame FUV becomes almost indistinguishable from the background as it is redshifted. \\
\textbf{Mrk 8}: This merging pair is very blue, with the two galaxies (three bright knots) visible in the DSS composite image. It is classified as a Wolf-Rayet galaxy, which suggests a very young burst \citep[4-6 Myr; see][and references therein]{est99}. \citeauthor{est99} find tidal tails indicated by low-intensity stellar structure in the deep V image. The FUV image does not resolve the separate galaxies, and the appearance stays similar for each redshift. This object is one of the high redshift star-forming galaxy analogs we have identified. We discuss the morphological analysis, leading to this result in the following sections (see \S 5.3-\S \ref{analogs}). \\ 
\textbf{NGC 520}: This galaxy is an extensively studied irregular galaxy with peculiar UV morphology. It is considered an intermediate-state merger between a gas-rich and gas-poor galaxy \citep{hib96}. In the DSS image in Fig. \ref{imgsb}(a), a dust lane is clearly visible. The infrared shows a bulge amongst the dust lane and the morphology of the dusty portion looks more like an edge-on galaxy \citep{sta90}. \cite{lotz06} artificially redshifted this galaxy and found it would be classified as a merger by G-$M_{20}$ at high redshift. We find that the high redshift simulations look very similar to tadpole galaxies that are observed in intermediate and high redshift deep HST images \citep{elm05,mel06}. The peculiar UV features, obscurred by the dust lane, do not appear in the high redshift images, and the brightest part is a small knot off-center of a diffuse extension, which is the nucleus of one of the merging pairs.\\ 
\textbf{NGC 1068}: This is a classic starburst AGN Seyfert 2 galaxy and has been the subject of numerous spiral galaxy studies for over a century. It is an infrared luminous spiral galaxy with tightly wound spiral arms that extend into fainter arms, forming a ring structure in the FUV. Most of the spiral structure is lost in the UV, leaving clumpy spots along the disk. It is another galaxy artificially redshifted by \cite{lotz06}. Its morphology is bulge-like due to the AGN. The arms and ring are lost in the high redshift images, and the nucleus becomes more spherically shaped at z$\sim4$. \\
\textbf{NGC 3079}: This edge-on galaxy is a Seyfert 2 galaxy. It has a bright bulge and is interacting with two other galaxies (NGC 3073 and MCG +09-17-009). The interacting system has extended HI only along NGC 3079 \citep{hib01}. It has mostly been studied for its properties of being a water MASER \citep[e.g.,][]{hen84,has85,kon06}. The redshifting process changes the appearance of ``clumpiness,'' so that in both the z$\sim1.5$ and 4 it looks like a chain galaxy with multiple knots and no bulge. We have also determined this to be a high redshift star-forming galaxy analog, based on the quantitative results described in the next sections (see \S 5.3-\S \ref{analogs}). \\
\textbf{NGC 3310}: This starburst spiral galaxy (Arp 217) appears irregular in the FUV. Two of the arms form loops in the upper and lower sides of the bulge; they can almost be resolved in the z$\sim1.5$ and 4 images. Even though it is a spiral like NGC 1068, it has a much younger population as can be seen from the color composite image. \cite{weh06} suggest that the accretion of smaller galaxies drives the star formation and evolution of the galaxy based on tidal debris remnants. \cite{bal81} propose that the thin loop feature is due to a collision with a dwarf galaxy. At z$\sim4$, NGC 3310 has nearly identical morphology to NGC 1068. The peculiar spiral structure is lost at high redshift.\\
\textbf{NGC 7673}: This is a Markarian starburst galaxy (Mrk 325) with a large blue clump, surrounded by smaller clumps to the left (see DSS image in Figure \ref{imgsb}). It has been studied in detail because of the irregular starburst clumps present in the system \citep[see][]{mar71,bor75,cas76,hom02,pas08}. HI is extended and the star formation is mostly within the inner optical region in the clumps \citep{hom02}. In the FUV, the galaxy looks highly compact with little detail resolved. As it is artificially redshifted, the galaxy continues to look more spherical. It is similar to Mrk 8 in this regard. We also have determined this to be a high redshift star-forming galaxy analog (see \S 5.3-\S \ref{analogs}). \\ 
\end{flushleft}

\subsection{GALFIT Models}
The images in Figure \ref{galfita} depict the 2D S\'ersic index GALFIT models with the residuals below. All FUV images show structure in the residual images. As the galaxies are redshifted, only a few bright knots remain in the residuals. The residual images help reveal the small features that might not be as apparent in the image of the object. It should be noted that the scales are different for each residual image. This was done to emphasize the remaining features. In most of the sample, these knots are the same in z$\sim1.5$ and 4. In particular, Mrk 8 has two knots that are the brightest parts of the two merging galaxies.  NGC 1068 reveals one of the tightly wound arms in z$\sim1.5$ and the nucleus in both z$\sim1.5$ and 4. NGC 3079 has three bright knots that are clear in each residual image. NGC 3310 and NGC 7673 reveal the faint, extended components that appear as background in the original images.

It was difficult to fit M 82 in the GALEX/FUV as a single galaxy, so we used three components, where one component covers the hot wind cones coming from the nucleus. The other two components fit along the nearly edge-on disk. The S\'ersic index values along the disk are very similar, so we averaged them for analysis.

\subsection{Morphologies}

One major aim of this paper is to determine whether nearby starburst galaxies from our sample are similar to high redshift star-forming galaxies in terms of structure. However, the morphology of star-forming galaxies has not been established. First of all, most of the work done in this topic is in the rest-frame UV, where clumpy, star-forming regions dominate the morphology. Only recently high resolution rest-frame optical became available, revealing the dynamical structure of these high redshift objects \citep[e.g.,][]{for06,law07b,gen08,epi09, for09}. The turbulent rotating star-forming disks suggested by \cite{gen08} agree with the clumpy morphology of the Ultra Deep Field galaxies described by previous authors \citep[e.g.,][]{elm05,elm07}. \cite{gen08} and \cite{elm07} argue that these clumpy, z$\sim$2, star-forming galaxies are in the process of collapsing into disks. \cite{epi09} find a high number of merging galaxies in a population of star-forming galaxies at redshifts $\sim$ 1.2 to 1.6. These multiple-clumpy objects often have S\'ersic values indicative of disturbed galaxies \citep[$n <$ 1;][]{elm07}. Based on this and the \cite{lotz06} and \cite{rav06} studies, we chose the morphological values typical for merging objects or multiple clumps ($M_{20}>-1.7$ and $n<0.8$ as the galaxy is redshifted) to select high redshift star-forming galaxy analogs from our nearby sample.

We are also exploring whether fits to high redshift galaxies could indicate bulge-like or disk-like structure, even though the galaxies have different structures in the local Universe \citep[see][]{lotz06,aki08,over08}. We start by comparing our morphological results with \cite{lotz06} and \cite{rav06}. Then, we describe the Figures \ref{morphplot}-\ref{galmorphb} and highlight properties of specific galaxies below.

\subsection{Comparisons of Derived Morphologies}
\cite{lotz06} find that $30\%$ of the GOODS LBGs (z$\sim4$; 11/36) and emission-line (z$\sim1.5$; 16/54) galaxies are bulge-like, based on $M_{20}<-1.6$ and G$\ge0.55$. We find, using the same criteria, that 20\% (11/54) of z$\sim1.5$ and 37\% (17/46) of z$\sim4$ galaxies are bulge-like. The majority of galaxies at z $\sim 1.5$ and 4 have intermediate/clumpy values of -1.6$<M_{20}<$-1.1, or values of $M_{20}>-1.1$ consistent with merger-like morphology. This result is expected for starburst galaxies at high redshift.

For the S\'ersic index, using $n>0.8$ as the region for exponential profiles (the same criteria as \cite{rav06}), we find that $\sim70\%$ have $n>0.8$ for z$\sim1.5$ and z$\sim4$. The majority of these have $M_{20}>-1.6$, implying multiple nuclei, but with an exponential profile from the S\'ersic index. Our results agree with \cite{rav06}, who found that $\sim70\%$ of z$\sim4$ LBGs have  $n>0.8$. However, for z$\sim1.5$, the authors find $\sim 40\%$ of starburst galaxies with $n>0.8$, differing significantly from our results. It is worth considering that the higher $n$ value in our sample might be due to S/N $<15$ for all objects\footnote{\cite{rav06} claim that $n$ is higher in GOODS than HUDF images for S/N$<15$.}. This can cause multiple clumps within a faint galaxy to be seen as separate objects as the galaxy background becomes indistinguishable from the sky background and leads to higher $n$. It may be that S\'ersic is better at picking out disk or exponential profile galaxies even with star-forming clumps present.

\subsection{Quantitative Morphologies}
The values for G, $M_{20}$ and $n$ are plotted in Figures \ref{morphplot}, \ref{galmorpha}, and \ref{galmorphb}. The shaded regions delineate systems that are merger- or bulge-dominated. These include the ranges $0>M_{20}>-1.1$ and $-1.7>M_{20}>-2.5$ \citep[see][]{lotz04}, respectively. We observe, in Figure \ref{morphplot}, that most of the high redshift sample fall within the intermediate and merger regions. This is not surprising since high redshift starburst galaxies are expected to experience frequent mergers. The regions, merger and bulge, are based on a small number (4) of local galaxies in \cite{lotz06}, and it is possible that these should be revised based on a larger sample. Our artificially redshifted objects range from mostly merger-dominated in the FUV to intermediate cases based on structures in both the \Bbnd \hspace{1pt} and \Vbnd+\ibnd bandpasses. Figures \ref{galmorpha} and \ref{galmorphb} display the plots of M$_{20}$ and G with $n$. In Figure \ref{galmorpha}, everything in the intermediate- and merger-dominated region and less than the $n=0.8$ line, should be considered an analog to high redshift clumpy star-forming galaxies, according to our criteria. 

Six of the 8 galaxies do not consistently stay in the same morphology region when artificially redshifted, due mostly to noise and the $(1+\rm{z})^{-4}$ surface brightness dependence. Only two objects remain in the same morphology region from z=0 to 4 in Figure \ref{morphplot}. NGC 1068 (E in Figures \ref{morphplot}-\ref{galmorphb}) stays in the bulge-dominated region, which is expected since it is a nearly face-on AGN spiral. NGC 3079 (F in Figures \ref{morphplot}-\ref{galmorphb}) stays in the merger-dominated region, which is due to the three distinct knots observed in the GALFIT residual images (see Figure \ref{galfita}). The comparison objects (see \S \ref{compgal}) for NGC 3079 in Figure \ref{comparison} reside in the merger-dominated area and may also be edge-on galaxies. NGC 3079 and its comparison galaxies are not mergers, and this discrepancy exposes a weakness in the technique. G-M$_{20}$ and $n$ values of edge-on galaxies might be misleading due to clumps along the disk. \cite{elm05} discuss the selection biases based on orientation, where the authors' definitions of clump cluster, doubles and chain galaxies might be similar objects viewed at different orientations. The chain galaxies in \cite{elm04} are determined to be face-on counterparts based on the clumpy distributions. While there are plenty of models and studies that agree with higher merger rates with redshift, assuming that all clumpy distributions are due to recent mergers is naive. \cite{gen08} show that the galaxies with large star-forming clumps tend to be turbulent. It is important to further analyze the clump structure as it relates to disk formation.

In Figure \ref{morphplot}, we observe that NGC 3310 (G in Figures \ref{morphplot}-\ref{galmorphb}) and NGC 7673 (H in Figures \ref{morphplot}-\ref{galmorphb}) move from the edge of the merger-dominated region at z=0. NGC 3310 moves into the bulge-dominated section at z$\sim1.5$ and back to the intermediate region at z$\sim4$. The residual image in Figure \ref{galfita} shows the change in the central bulge for each redshift range. NGC 7673 stays within the intermediate region, shifting slightly in M$_{20}$, and the residual image shows that the two bright, off-center knots appear in each redshift. Mrk 8 (C) is an object that moves from a merger-dominated system to well within the intermediate region as it is redshifted. In the residual, the FUV has four bright knots that slowly disappear as it is redshifted, leaving two bright knots close to the geometric center. Arp 269 (A, A*) is peculiar and a special case since it ``merges'' into one object at z$\sim4$. The residual images reveal that the two galaxies become very smooth compared with the numerous starburst clumps in the FUV, leaving little residual structure. Observing this type of change in morphology is helpful in exploring systems that we know to have disturbed morphologies nearby and are quantified as disk-like when redshifted, revealing biases due to the ``smoothing'' of features at high redshift.

In the $n$-$M_{20}$ and G plots (Figures \ref{galmorpha} and \ref{galmorphb}), we note that most of our local sample is below the $n=0.8$ line. NGC 1068 (E) is, again, consistently in the bulge-dominated area as it is redshifted. Although, it is important to mention that NGC 1068 moves progressively from a strong exponential profile toward the border of the $n=0.8$/M$_{20}=-1.7$ line as it is redshifted. This is most likely due to the profile obtaining a steep cut-off with the applied size evolution as it is redshifted to z$=4$ (see \S \ref{artred}). NGC 1068 is a good standard for testing the morphology of exponential disk and bulge types at different redshifts. 

M 82 is also a nearly edge-on galaxy, and, as discussed for NGC 3079, this presents issues that should be taken into consideration. We use two different wavelength images to artificially redshift M 82. We use GALEX/FUV and ACS \Bbnd-band images for z$\sim0$. The G, $M_{20}$ and $n$ morphologies are similar for z$\sim0$ and z$\sim1.5$. This $n$ value is questionable, due to the wind cones in the FUV that dominate and were difficult to fit in a 2D profile. It is also notable that the wind creating the FUV features for M 82 presumably occurs in galaxies at high redshift. In this case, we would be measuring the structure of the wind and not that of the host galaxy at high redshifts.

\subsection{Structural Analogs to Clumpy Star-forming Galaxies \label{analogs}}
We have determined that 3 of the 8 GALEX/FUV z$\sim0$ objects are high redshift clumpy star-forming galaxy analogs, because of their locations in the G-$M_{20}$ and $n$-$M_{20}$ plots after they are redshifted. We describe their properties based on Figures \ref{morphplot}-\ref{galmorphb}. We focus on those galaxies which have clumpy or merger-like structures to define the region in parameter space which distinguishes the analogs to high redshift star-forming galaxies.

Mrk 8 (C in Figures \ref{morphplot}-\ref{galmorphb}) is consistently in the merger region for $n$, and merger to intermediate region for $M_{20}$, making it an analog to clumpy star-forming galaxy at high redshift. After it is redshifted to z$\sim1.5$, it moves into the intermediate area, with a smoother G profile, but it is still within the LBG population. Mrk 8 is a two galaxy system in the process of merging. At z$\sim1.5$, it is in the merger area for z$\sim1.5$ and slips into the intermediate region at z$\sim4$. It stays below n$=0.8$ in the S\'ersic plot, which indicates the merger/disturbed profile. 

NGC 3079 (F in Figures \ref{morphplot}-\ref{galmorphb}) is in the merger region for G and $M_{20}$ for all redshifts. It changes S\'ersic morphology from an exponential disk at z$\sim0$ to a merger at $z\sim1.5$ and 4. The three bright knots, discussed in the previous section, are what drive the G-$M_{20}$ classification, and $n$ becomes shallower due to the loss of the fringe light distribution in the background. This is an edge-on galaxy and not typical for a clumpy star-forming galaxy structure (when viewing the nearby GALEX/FUV). Therefore, it is an example of how orientation and loss of low surface brightness features can play a role in  producing objects that are similar to high redshift clumpy objects.

NGC 7673 (H in Figures \ref{morphplot}-\ref{galmorphb}) remains in the $n$ merger and G-$M_{20}$ intermediate area from z$\sim 0$ to 4. NGC 7673 is a peculiar galaxy that has strong FUV emission and a fairly smooth G and $M_{20}$ distribution of light. The two bright knots in the residual images (see Figure \ref{galfita}) are large star-forming clumps, which get treated as multiple nuclei. This is a similar galaxy as Mrk 8 in terms of having bright star-forming clumps analogous to those observed in high redshift star-forming galaxies. 

\subsection{Comparison Galaxies \label{compgal}}

We also inverted the comparison process by selecting six galaxies from \cite{voy} and \cite{lotz06} GOODs samples that are structurally similar to our scaled and redshifted versions of Mrk 8, NGC 3079, and NGC 7673. These comparisons are based on similarity of the measured G, $M_{20}$ and $n$ values, as well as visual inspection. The point of this exercise is to check whether our artificially redshifted local galaxies resemble specific high redshift objects. Figure \ref{comparison} shows the six galaxies, two for each high redshift star-forming galaxy analog (from left to right, Mrk 8, NGC 3079, and NGC 7673). The top (bottom) row displays z$\sim1.5$ \Bbnd-band (z$\sim4$ \Vbnd+\ibnd-band) images. We list the morphological parameters G, $M_{20}$, $n$, Petrosian radius (r\_p), and the nearby galaxy name on the image. All objects are from the \cite{lotz06} sample, except for the NGC 3079 structural analog at z$\sim1.5$, which is from \cite{voy}.

The Mrk 8 (left panels) and NGC 7673 (right panels) comparison objects are compact multiple clump objects. The residuals from the $n$ model (not shown) reveal the multiple knots similar to Mrk 8 and NGC 7673 (see Figure \ref{galfita}). The comparison objects for NGC 3079 (middle panels) also have distinct, multiple knots in their residual images (not shown) and look like edge-on galaxies. Their $n$ values may have them classified as mergers, when these are most likely star-forming clumps along the disk of an edge-on galaxy.

%%%%%%%%%%%%%%%%%%%%%%%%%%Summary%%%%%%%%%%%%%%%%%%%%%%%%%%%%
\section{Summary\label{summary}}
We use our GALEX/FUV sample of nearby galaxies to simulate galaxies in the GOODS/ ACS \Bbnd- and \Vbnd+\ibnd-bands and determine their quantitative structural properties to compare with distant (z$\sim$ 1.5 and 4) star-forming galaxies from \cite{voy} and \cite{lotz06}. We investigate how structures of high redshift starburst galaxies compare with structures of nearby galaxies that are artificially redshifted with the proper luminosity and size scaling. We find some local starbursts have the structural features of high redshift star-forming galaxies with disturbed or merging profiles ($n<0.8$ and $M_{20}>-1.7$) when evolved in star formation rate and size. This is carried out by manipulating FUV-to-optical images of the nearby system and comparing these to distant galaxies observed in the GOODS fields. We also demonstrate that, due to the effects of ``smoothing'' and limited S/N, disk-like or bulge-like structural parameters derived from images do not uniquely indicate the presence of these features in the faint, distant galaxies. Our analysis highlights the importance of combining the 2D profile with nonparametric methods: it can identify analogs to disturbed or merging high redshift star-forming galaxies, where the G and $M_{20}$ might not be as distinguishable. 

Our main results are:

\begin{enumerate}
\item We find three objects whose structures resemble those of interacting, or clumpy star-forming galaxies from z$\sim1.5$ to $4$: Mrk 8, NGC 3079 and NGC 7673. Nearby galaxies similar to these are worth studying in more detail to help determine the structures of high redshift galaxies, such as LBGs.
\item The morphology indicators for many of our artificially redshifted sample vary significantly with redshift. As each galaxy is redshifted, they move from one type to another. Care is required in applying these results from quantitative structural measurements in terms of physical components, such as exponential stellar disk distortions due to mergers, or the presence of bulges. A merging, clumpy nearby galaxy may have the structural indexes of an exponential disk at high redshift simply because of surface brightness dimming and loss of resolution.
\item \cite{rav06} found that GOODS galaxies observed with S/N $<$ 15 tend to have higher S\'ersic $n$ values. We tested this effect by placing nearby galaxies, of varying morphologies, and artificially moving them to high redshifts. The S\'ersic index varies with redshift in a manner consistent with that found by \citeauthor{rav06} The residuals from model fits to the images show that this is a result of the confusion of lower surface brightness regions with the sky noise at higher redshift.
\item Based on currently available data, it is difficult to determine what types of interactions occur at high redshift using G, $M_{20}$ and $n$, although it is possible with residual 2D images to get a closer look at the dominating surface features in the fits. This is perhaps a problem where multiple wavelength analysis can provide a more cohesive, overarching view of morphologies. 
\item It is clear that a larger sample of FUV observations of local galaxies is needed to classify galaxies using the $M_{20}$, G and $n$ relations. Using multiple rest-frame wavelengths will help to distinguish the biases due to color by cross-matching samples. Using the HST/NICMOS (Near Infrared Camera and Multi-Object Spectrometer) or other near-infrared (NIR) observations with this method would shed light on many of the above problems, since the rest-frame wavelengths would be in the optical, although we then face the problem of reduced angular resolution. Surveys performed by the James Webb Space Telescope (JWST) \citep{gar06} will allow rest-frame optical structures to be determined for high redshift objects, as will studies with 8-10 m class telescopes from the ground with adaptive optics.
\end{enumerate}

The results from these types of studies, as well as comparisons with larger samples, will test our assumption that some high redshift star-forming galaxies are scaled versions of nearby objects.

\acknowledgments We thank the anonymous referee for the helpful comments, which improved this paper. We are grateful to S. Ravindranath for discussions concerning luminosity evolution and for the GALFIT tutorial. We also thank B. Holwerda for conversations about SExtractor and morphological analysis, and C. Peng for assistance with GALFIT. SMP was funded by NASA's NNX07AJ95G. DFdM was funded by HST GO-10776.13-A, JWST NNX07AR82G and NASA NNX07AJ95G. JSG's research on starburst galaxies was supported in part by the National Science Foundation through grant AST0708967 to the University of Wisconsin-Madison. 

The GALEX and GOODS data in this paper were obtained from MAST at the Space Telescope Science Institute (STScI). STScI is operated by the Association of Universities for Research in Astronomy, Inc., under NASA contract NAS5-26555. Support for MAST for non-HST data is provided by the NASA Office of Space Science via grant NAG5-7584 and by other grants and contracts. DSS data were obtained using the Aladin interactive sky atlas \citep{bon00}. The Digitized Sky Survey was produced at the Space Telescope Science Institute under U.S. Government grant NAG W-2166. SDSS data were obtained from the SDSS archive Data Release Six. Funding for the SDSS and SDSS-II has been provided by the Alfred P. Sloan Foundation, the Participating Institutions, the National Science Foundation, the U.S. Department of Energy, the National Aeronautics and Space Administration, the Japanese Monbukagakusho, the Max Planck Society, and the Higher Education Funding Council for England.

\mybib 
\imgsa
\clearpage
\imgsb
\clearpage
\galfita
\clearpage
\morphplot 
\clearpage
\morpha
\clearpage
\morphb 
\clearpage
\comparison
\clearpage
\gentable
\clearpage
\fuvtable

\end{document}